# Optomechanical self-channelling of light in a suspended planar dual-nanoweb waveguide

A. Butsch[1], C. Conti[2], F. Biancalana[1] and P. St. J. Russell[1]

[1]Max Planck Institute for the Science of Light,

Guenther-Scharowsky-Str. 1, 91058 Erlangen, Germany

[2]Dept. Mol. Med. and CNR-ISC, Dept. of Physics, University La Sapienza,

Piazzale A. Moro 2, 00185 Rome, Italy

It is shown that optomechanical forces can cause nonlinear self-channelling of light in a planar dual-slab waveguide. A system of two parallel silica nanowebs, spaced ~100 nm and supported inside a fibre capillary, is studied theoretically and an iterative scheme developed to analyse its nonlinear optomechanical properties. Steady-state field distributions and mechanical deformation profiles are obtained, demonstrating that self-channelling is possible in realistic structures at launched powers as low as a few mW. The differential optical nonlinearity of the self-channelled mode can be as much as ten million times higher than the corresponding electronic Kerr nonlinearity. It is also intrinsically broadband, does not utilize resonant effects, can be viewed as a consequence of the extreme nonlocality of the mechanical response, and in fact is a notable example of a so-called "accessible" soliton.

Attractive and repulsive optical forces can appear between two coupled waveguides when light is launched into them [1-4]. Recently, experimental observations of transverse gradient forces and mechanical bending have been reported for a nano-structured waveguide beam [5], stacked ring microcavities [6], a periodically patterned "zipper" cavity [7] and coupled waveguide beams [8]. Optical forces can cause transverse deformations in coupled waveguide systems, which in turn change the effective refractive index, giving rise to the so-called "mechanical Kerr effect" [9, 10].

In this Letter, we present a novel optomechanical phenomenon - the self-channelling of light, which we study theoretically in a system of two glass waveguides ("nanowebs") suspended between the walls of a capillary fiber. When an optical mode is launched into the nanowebs, optical gradient forces cause them to bend inwards or outwards depending on the symmetry of the optical mode profile perpendicular to the nanoweb plane [1]. This mechanical deflection increases the effective refractive index of the mode, causing self-focusing of light within the nanoweb plane and altering the radiation-induced pressure distribution. We show that this can lead to stable self-channelling at power levels as low as a few mW in experimentally realistic structures [11]. The refractive index of the self-channelled mode turns out to be highly sensitive to small changes in the optical power, indeed the resulting differential optomechanical nonlinear coefficient can exceed the $\gamma$-coefficient in conventional fibers by up to seven orders of magnitude [12]. The magnitude of the optomechanical nonlinearity also remains high even when the nanowebs have quite dissimilar thicknesses.

Since a point force applied at one position causes a deflection that extends over the whole nanoweb, the optomechanical elastic response is highly nonlocal – a characteristic that is known to favour stable self-trapping [13]. Indeed, self-channelled modes in systems with a nonlocal response are an example of spatial "accessible" solitons [14]. A temporal analogue of such "accessible" solitons was recently proposed in [15].

Fig. 1 shows the generic structure under study: the $z$-axis points along the fibre axis and the $y$-axis is perpendicular to the nanoweb plane. The webs have width $L$, thickness $w$ and the air gap width is $h$. The structure supports both TE and TM polarized modes.

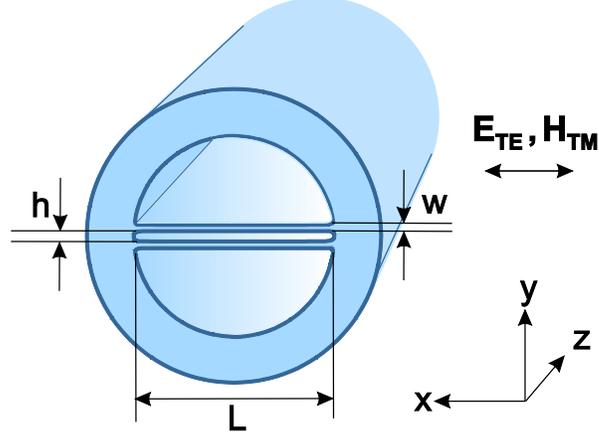

FIG. 1: Sketch of the dual-nanoweb fibre structure. Light propagates in the $z$-direction. The electric field of the TE mode and the magnetic field of the TM mode point parallel to the $x$-axis.

For the general case of an asymmetric dual waveguide we define a parameter $s$, so that the web thicknesses become $w+s$ and $w-s$. In this case the dispersion relation can be written in the following implicit analytical form:

$$pw = \arctan\left(\frac{A(1+\tanh gh) \pm \sqrt{(1+A^2-B^4\cos^2 2ps)\tanh^2 gh + A^2 + A^4}}{A^2 - (1+B^2\cos 2ps)\tanh gh}\right) - m\pi \quad (1)$$

where $p = k_0\sqrt{n_g^2 - n_{\text{eff}}^2}$ and $g = k_0\sqrt{n_{\text{eff}}^2 - 1}$ are the wavevector components in dielectric slabs with refractive index $n_g$ and in air, respectively, and $A$ and $B$ are given by $A = 2gp\xi/(g^2 - p^2\xi^2)$ and $B = (g^2 + p^2\xi^2)/(g^2 - p^2\xi^2)$. The parameter $\xi = 1$ for TE and $1/n_g^2$ for TM polarization, allowing both polarizations to be conveniently treated in a single analysis. For each value of integer $m \geq 0$ two modes exist, one (+ sign in Eq. (1)) corresponding to a mode with $2m$ field nodes and the other (− sign) to a mode with $2m+1$ nodes. For $s = 0$ Eq. (1) reduces to expressions for the symmetric and anti-symmetric modes of a dual-slab waveguide derived in [3].

The electromagnetic force acting on each slab can be calculated by means of the Maxwell stress tensor. Since it is proportional to the intensity of light in the waveguide mode, the field amplitudes must be related to the optical power through the flux of the $z$-component of the time averaged Poynting vector $S_z$. The total optical power $P$ and the power density $p$ per unit length in $x$-direction are calculated as follows:

$$P = \int_{-\infty}^{\infty}\int_{-\infty}^{\infty} \overline{S}_z dxdy, \qquad p = \int_{-\infty}^{\infty} \overline{S}_z dy \qquad (2)$$

The optical gradient pressure is obtained by evaluating the time-averaged y-component of the Maxwell stress tensor:

$$<T_{yy}> = \varepsilon_0 (Q_y - Q_x - Q_z)/4, \qquad Q_i = |E_i|^2 + c^2 \mu_0^2 |H_i|^2. \qquad (3)$$

The pressure-induced deflection $\delta(x)$ of a slab (thickness w), pinned rigidly at each edge ($x = \pm L/2$) to a solid glass wall, is found using the standard plate deflection equation:

$$d^4\delta(x)/dx^4 = 12(1-v^2)\sigma(x)/(Ew^3), \qquad (4)$$

where $E$ is Young's modulus, $v$ Poisson's ratio and $\sigma(x)$ the pressure distribution. The boundary conditions at $x = \pm L/2$ are $\delta = 0$ and $d\delta/dx = 0$.

Since the nanowebs are much wider than they are thick and the variation of the local modal index $n(x)$ is slow, the x-dependent modal distribution $e(x)$ of the deflected structure can be accurately found by numerically solving the Helmholtz equation:

$$d^2 e(x)/dx^2 + k_0^2 (n^2(x) - n_m^2) e(x) = 0. \qquad (5)$$

Making use of these connections between radiation-induced pressure, mechanical deformation and transverse electromagnetic field distribution, an iterative cycle of numerical calculations can be established to seek self-consistent self-channelled solutions. Starting with a trial pressure profile $\sigma_0(x)$, we calculate the deflection $\delta(x)$ using (4). By solving (5) we obtain the mode index $n_m$ and the transverse field distribution $e(x)$. A new pressure profile $\sigma_1(x)$ is then calculated using (3) and the cycle restarted. Keeping the total optical power $P$ constant, we iterate this procedure several times until it converges to a self-consistent steady-state solution.

We now consider a structure with $w = 200$ nm, $h = 300$ nm and $L = 70$ μm, made from fused silica ($n_g = 1.45$). Such structures can be produced by fibre drawing [11, 16]. Fig. 2(a) and (b) shows the self-channelling of the $m = 0$ even and odd TE modes. The wavelength of the light is 800 nm, Young's modulus 72.5 kN/mm$^2$, Poisson's ratio 0.17 and the optical power was set to 100 mW. Self-channelling causes the nanowebs to be attracted for even modes and repelled for odd modes. The maximum deflection amplitude is 1.4 nm for the even mode (webs bent inward) and 2.5 nm for the odd mode (webs pushed outward). These relatively small deflections are sufficient to create a guiding index profile in the x-direction.

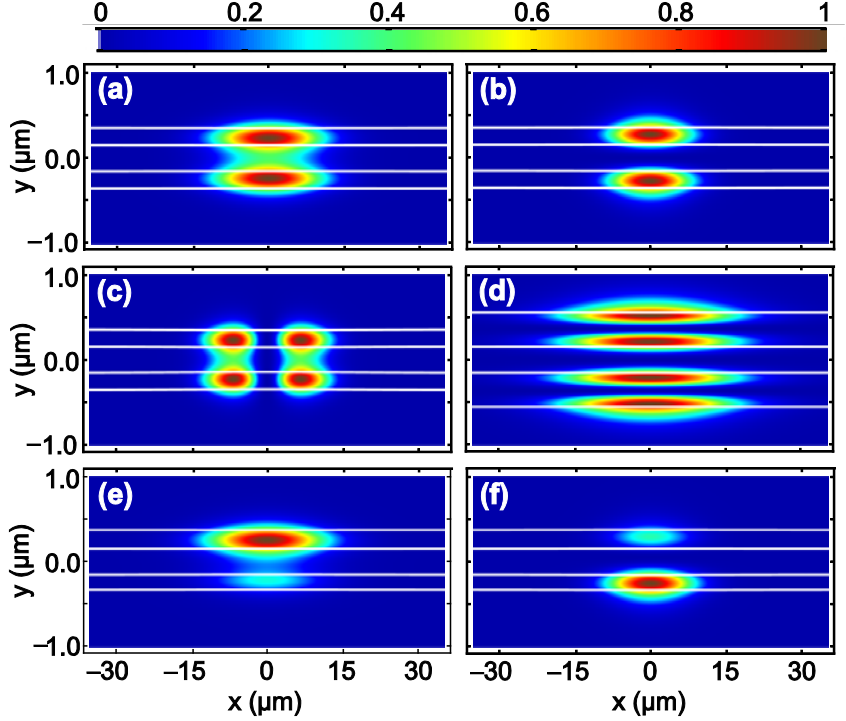

FIG. 2: Normalized Poynting vector distributions of selected TE self-channelled modes at a power of 100 mW. (a) Even $m = 0$ mode and (b) odd $m = 0$ mode for $w = 200$ nm, $h = 300$ nm, $L = 70$ μm, $\lambda = 800$ nm, $P = 100$ mW. (c) Higher order self-channelled TE mode with two lobes in the x-direction ($w = 200$ nm). (d) Higher order self-channelled TE mode with two lobes in each nanoweb in the y-direction ($w = 400$ nm, $h = 300$ nm, $\lambda = 600$ nm). (e) Even and (f) odd $m = 0$ mode for the asymmetric structure with $w_1 = 220$ nm and $w_2 = 180$ nm, the upper web being thicker.

Note that when the air gap gets smaller, the index increases for even modes and reduces for odd modes. Thus both modes can be localized (channelled) in the central region of the waveguide. For the parameters chosen, the lateral extent of the $m = 0$ TE mode is ~20 μm (even) and ~16 μm (odd). Higher-order self-channelled modes can also exist (Fig. 2(c)). Changing the parameters to $w = 400$ nm, $h = 300$ nm and $\lambda = 600$ nm, a self-channelled $m = 1$ mode appears with two lobes across each nanoweb (Fig. 2(d)).

It is interesting to enquire how sensitive the self-channelling is to fabrication imperfections, such as unequal nanoweb thicknesses. In Fig 2(e&f) the mode profiles are plotted for a structure with web thicknesses 220 and 180 nm. It can be seen that the self-channelling is robust against quite large degrees of asymmetry.

The power dependence of the modal index for the $m = 0$ even and odd self-channelled TE modes ($w = 200$ nm, $h = 300$ nm) is presented in Fig. 3 for three different values of $L$. The lower limit of $P$ in the calculations is determined by the need for the self-channelled mode to be confined tightly enough in the x-direction so as not to be affected by the presence of the supporting capillary. By appropriate choice of parameters, powers as small as 10 mW can be sufficient to achieve optomechanical self-channelling in a realistic structure.

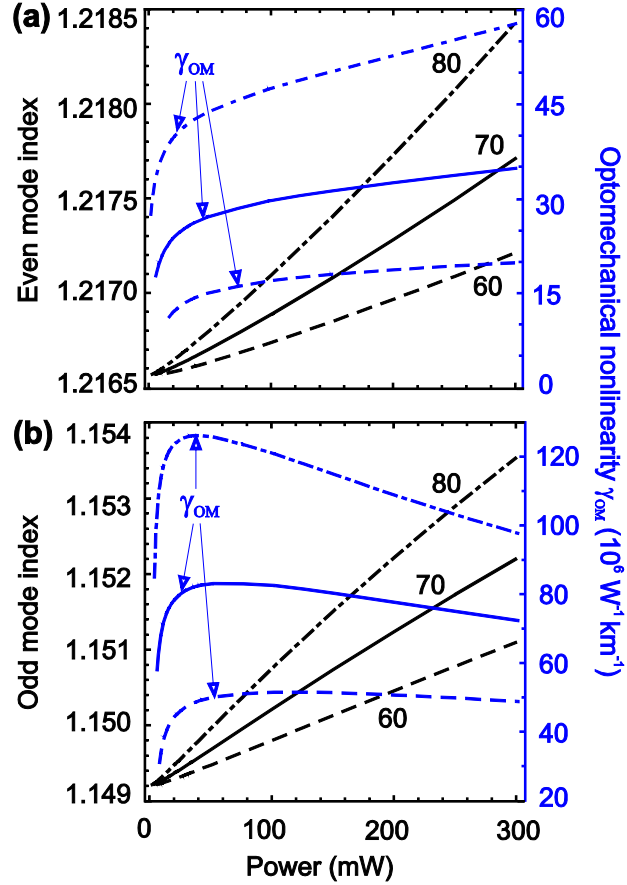

FIG. 3: Power dependence of the modal index and the differential optomechanical nonlinearity. The modal indices converge to the same value at zero power. (a) Even $m = 0$ TE mode in a structure with $w = 200$ nm, $h = 300$ nm, $\lambda = 800$ nm and three different values of width $L$: 60 µm (dashed), 70 µm (solid) and 80 µm (dash-dotted line). (b) the same as (a) for the odd $m = 0$ TE mode.

For both even and odd modes the modal index increases monotonically with power, the slope being steeper for wider, i.e., mechanically more compliant, structures. The differential change of modal index with variations in power can be regarded as a measure of nonlinearity of the system. We therefore define a differential optomechanical nonlinearity $\gamma_{om}$:

$$\gamma_{om} = k_0 \partial n_m / \partial P \quad \text{W}^{-1}\text{km}^{-1}, \tag{6}$$

and plot its value as a function of the optical bias power in Fig. 3(a) for the even and (b) for the odd $m = 0$ TE modes.

The intrinsic (i.e., electronic) $\gamma$-coefficient can be calculated from the nonlinear refractive index of silica and the effective nonlinear mode area [17]. For the dual-web structure analyzed above it is ~1 $W^{-1}km^{-1}$, lying between the value typical of a small-core photonic crystal fibre (~$0.2\times10^3$ $W^{-1}km^{-1}$) and that of a hollow-core PCF (~$2\times10^{-2}$ $W^{-1}km^{-1}$) [18]. With peak values of order ~$10^8$ $W^{-1}km^{-1}$, $\gamma_{om}$ is more than ten million times larger than the Kerr nonlinearity.

Although asymmetry causes dephasing and weakens the interaction between the waveguides, it turns out that even when it is strong (e.g., $w_1 = 250$ nm, $w_2 = 150$ nm) the nonlinear coefficient $\gamma_{om}$ remains very high, dropping to roughly one third its value in the symmetric structure for the even and to two thirds for the odd $m = 0$ TE mode, which means that fabrication tolerances for these structures can be quite relaxed.

The even and odd modes exhibit different types of self-channelling behaviour. In Fig. 4 $\gamma_{om}$ is plotted as a function of $h$ and wavelength for $L = 70$ µm, $w = 200$ nm and $P = 300$ mW. The maximum value of the nonlinearity for the given range of parameters reaches ~$10^9$ $W^{-1}km^{-1}$ for the even $m = 0$ TE mode (Fig. 4 (a)) and ~$5\times10^8$ $W^{-1}km^{-1}$ for the odd mode (Fig. 4 (b)). Furthermore, the optical gradient force increases as the web separation decreases [1]. The $m = 0$ odd mode, however, cuts off when $h$ falls below a certain value, for example at $w = 200$ nm and $\lambda = 1.3$ µm the odd TE mode disappears at $h = $ ~250 nm. This cut-off spacing decreases with wavelength, as indicated by the red curve in Fig. 4 (b). A structure in which only the even mode is self-channelled is potentially very interesting experimentally, since the system will become single-mode, making measurements much more straightforward. Due to weaker confinement in the y-direction, the odd mode shows maximum nonlinearity at longer wavelength compared to the even mode. Note that at shorter wavelengths the modes are more tightly confined in the y-direction, causing the nonlinear coefficient to fall off more rapidly with increasing $h$.

The TM modes show qualitatively similar behaviour, except that at very small spacings ($h < 100$ nm) the $m = 0$ even TM mode experiences higher radiation pressure than the even TE mode [3]. When the distance between the webs is decreased further, or alternatively the optical bias power is increased, the webs will touch (in case of even modes), causing a discontinuity in the optomechanical nonlinearity.

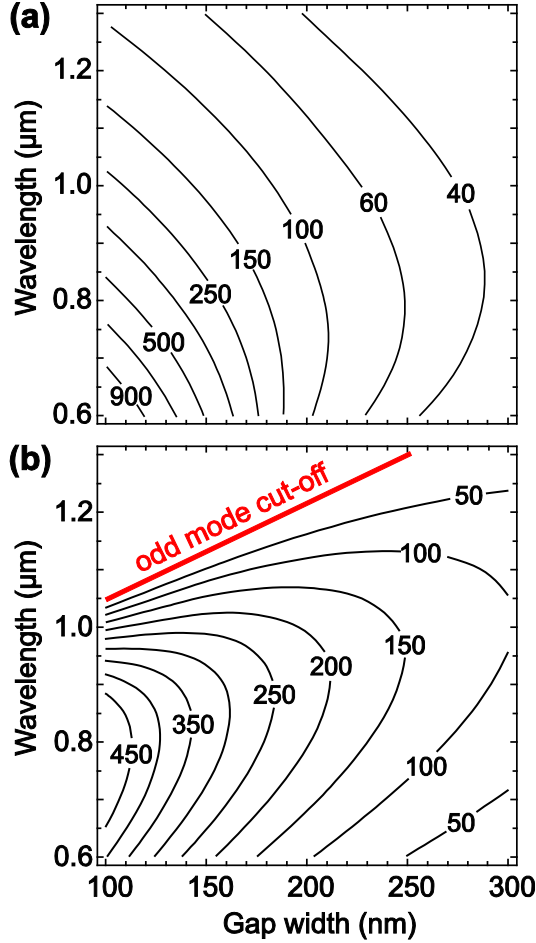

FIG. 4: Contours of constant optomechanical nonlinear coefficient $\gamma_{om}$ as a function of gap width $h$ and wavelength (in units of $10^6\,\text{W}^{-1}\text{km}^{-1}$). The nonlinear coefficient based on the Kerr-effect is only $\sim 1\,\text{W}^{-1}\text{km}^{-1}$. (a) Nonlinearity of the even $m = 0$ TE mode of the dual-nanoweb waveguide with web thickness $w = 200$ nm, width $L = 70$ µm and optical bias power $P = 300$ mW. (b) nonlinearity of the odd $m = 0$ TE mode, red curve showing the cut-off condition for the odd mode at given parameters.

As mentioned above, the stability of the self-channelled modes and the very large differential nonlinear coefficient is a consequence of the extreme nonlocality of the mechanical response. Although the self-channelling nonlinearity is much stronger than the electronic Kerr-nonlinearity of the glass, it is, of course, much slower. The time constant of the optomechanical nonlinear response is limited by the mechanical resonant frequencies of the dual-nanoweb (sub-MHz range, response time ~10 µs), which are however considerably faster than in other nonlocal media such as liquid crystals (~1 ms, [19]) and photorefractive materials (~1 s, [20]). In addition, the nonlinear response can be further enhanced by modulating the light at the acoustic resonant frequency, as suggested in several other contexts [9, 21, 22].

The main purpose of this Letter has been to show for the first time that steady-state self-channelled modes can exist in this dual-nanoweb system. We now briefly indicate how the analysis can be extended to describe the dynamics of the self-channelling process. A paraxial model can be constructed for the propagation of an optical beam with intensity $|A(x, z)|^2$:

$$2ik\partial_z A + \partial_{xx} A + 2k^2(\Delta n / n_{\text{eff}})A = 0 \qquad (7)$$

where $k = \omega n_{\text{eff}}/c$, $n_{\text{eff}}$ being the effective modal index of the undistorted nanoweb pair and $\Delta n = n(x) - n_{\text{eff}}$ the perturbation due to optomechanical forces. At equilibrium, for a given power $P$, the stress-induced deformation can be approximated by a parabolic function, leading to $\Delta n(x) \approx n_0(P) - n_2(P) x^2$. Explicit expressions for $n_0(P)$ and $n_2(P)$ can then be obtained from a second order expansion in $x$ of the Green's function of Eq. (4). This approach predicts the existence of a whole family of Hermite-Gaussian solutions, including fundamental and higher-order spatial solitons. A detailed study of the dynamics of nonlocal optomechanical solitons (a unique realization of the so-called "accessible" solitons [14]) will be reported in a future publication.

In conclusion, dual-nanoweb fibers are a promising vehicle for attaining ultra-high nonlinear coefficients in optical fibers. In contrast to previously reported nonlocal nonlinear systems [19, 20], they allow the formation of stable self-trapped modes that can propagate over many metres along a flexible pathway. This may permit the observation of spatiotemporal solitons (light bullets), and related phenomena such as supercontinuum generation inside spatially self-trapped beams.

**Acknowledgements**

C. Conti acknowledges support from ERC grant 201766.